\documentclass[12pt,a4paper]{article}

\usepackage{graphicx}

\begin{document}
\begin{flushright}\normalsize
TAUP-2524-98\\
M/C-TH-98/15
\end{flushright}
\begin{center}
{\bf\Large Two Photon Reactions at High Energies}
\end{center}
\vskip 1truecm
\begin{center}
A.~Donnachie\\
Department of Physics and Astronomy\\
University of Manchester, Manchester M13 9PL, United Kingdom\\
email: {\tt ad@a3.ph.man.ac.uk}
\end{center}
\begin{center}
H.G.~Dosch\\
Institut f\"ur Theoretische Physik der Universit\"at Heidelberg\\
Philosophenweg 16, D-69120 Heidelberg, Germany\\
email: {\tt h.g.dosch@thphys.uni-heidelberg.de}
\end{center}
\begin{center}
M.~Rueter\footnote{Supported by a MINERVA-fellowship}\\
School of Physics and Astronomy\\
Department of High Energy Physics, Tel Aviv University\\
69978 Tel Aviv, Israel\\
email: {\tt rueter@post.tau.ac.il}
\end{center}
\vskip 2truecm
\begin{center}
{\bf Abstract}
\end{center}
Cross sections for the reactions $\gamma^{(*)}\gamma^{(*)}
\rightarrow$ hadrons and $\gamma^{(*)}\gamma^{(*)} \rightarrow$ 2
vector mesons are calculated as functions of energy ($\surd s \ge 20$
GeV) and photon virtualities. Good agreement with experiment is
obtained for the total hadronic cross section and, after allowing for
a valence-quark contribution from the hadronic part of the photon,
with the photon structure function at small $x$. The cross section for
vector meson production are shown to be experimentally accessible for
moderate values of $Q^2$. This is sufficient to probe the nature of
the hard pomeron which has recently been proposed.  \vfill \eject

\section{Introduction}

In the colour-dipole picture of high-energy scattering the scattering
amplitude is expressed as a superposition of dipole-proton
\cite{Nik91} or dipole-dipole \cite{KraD91} amplitudes. In the model
of the stochastic vacuum \cite{Dos87,DosS88} as applied to high-energy
scattering \cite{KraD91,Nac91,DFK94,BerN98} the dipole-dipole
scattering amplitude is realized as the vacuum expectation value of
two Wegner-Wilson loops. Any hadronic (or photonic) scattering
amplitude can be expressed in terms of these amplitudes and the
transverse wave functions of the hadrons (or photons) involved. Thus
the model correlates a wide range of phenomena: hadron-hadron
scattering \cite{DFK94,BerN98}, vector meson electro- and
photoproduction \cite{DGKP97,KDP98} and deep inelastic scattering
\cite{Rue98}. This is of considerable relevance as understanding the
details of high-energy scattering at the microscopic level requires
the investigation of many different processes within the context of a
unified description.

The relevance of the dipole approach to deep inelastic scattering was
first stressed by Nikolaev and Zakharov \cite{Nik91} and subsequently
elaborated in a number of papers \cite{Nikx}. More recently the
dipole-proton cross section $\sigma_T(s,R_D)$ has been obtained from
electroproduction data \cite{nem} as a function of energy and the
dipole radius $R_D$. The data span dipole radii from 0.2 to 1.1 fm,
and the cross section changes by more than an order of magnitude
across this range. The model of the stochastic vacuum applied to high
energy scattering is in reasonable accord with these results
\cite{dr}.

In this paper we investigate high-energy $\gamma^{(*)}\gamma^{(*)}$
reactions. We calculate and predict the total real $\gamma\gamma$
cross section, the real photon structure function at small $x$ and
reactions of the type
\begin{equation}
\gamma^{(*)}\gamma^{(*)} \rightarrow V_{1}V_{2},
\end{equation} 
where $V_{1}$ and $V_{2}$ can be any two vector mesons. These may be
light quarkonia i.e. $\rho$, $\omega$, $\phi$ or heavy quarkonia for
which we restrict consideration to $J/\psi$.

The physics interest in these latter processes is that they are the
closest one can get to the ideal of onium-onium scattering
\cite{muel}. They provide as direct a measurement as is possible of
colour dipole-dipole scattering, and the size of the dipoles can be
tuned by the choice of vector meson or by the photon virtuality or
both. The $\gamma^{(*)}\gamma^{(*)}$ reactions have a considerable
advantage over $\gamma^{(*)} p$ reactions as the dipole scattering is
measured directly and the complications arising from a proton target
are avoided.

The ultimate goal is that the relevant experiments are performed to
confirm the dipole approach and hence to obtain the dipole-dipole
cross sections $\sigma_T(s,R_{D_1},R_{D_2})$ independently of specific
theoretical models for a range of dipole radii $R_{D_1}$, $R_{D_2}$.
At present one has to make use of specific models to provide an
estimate of the cross sections. Our calculations demonstrate that
significant measurements can be made with exisiting energies and
luminosities. The present calculations are based on the fusion of two
models. The absolute size and the dependence of the
$\gamma^{(*)}\gamma^{(*)}$ cross sections on the photon virtuality
$Q^2$, and where relevant the choice of vector meson, are given by the
model of the stochastic vacuum \cite{Dos87,DosS88} applied to
high-energy scattering \cite{Nac91,DFK94} at a center of mass energy
of about 20 GeV. This model can say nothing about the energy
dependence, which is taken to be given by a two-pomeron model proposed
recently \cite{DL98} and which has subsequently been applied to the
model of the stochastic vacuum \cite{Rue98}.

The relevant aspects of these models are discussed in Section 2. In
Section 3 we show that the model is in good agreement with data on the
total hadronic $\gamma\gamma$ cross section $\sigma_{\gamma\gamma}$,
and we make predictions for the photon structure function
$F_{2}^{\gamma}(x,Q^2)$ at very small $x$ and over a wide range of
$Q^2$. At the higher values of $x$ where data exist, combining our
model with the VMD contribution corresponding to non-pomeron Regge
exchange provides fair agreement with these data. Section 4 deals with
the $\gamma^{(*)}\gamma^{(*)} \rightarrow V_1V_2$ cross sections in
detail, and conclusions are drawn in Section 5.

\section{The Model}

The model of the stochastic vacuum (MSV) \cite{Dos87,DosS88} is based
on the asumption that the infrared behaviour of QCD can be
approximated by a Gaussian (i.e. factorizable) stochastic process in
the gluon field strength tensor. With this simple assumption one
obtains in non-Abelian gauge theories the area law for the
Wegner-Wilson loop and hence confinement. Applying the model to the
formalism for quark-quark scattering in the fempto-universe developed
in \cite{Nac91} one can express any hadronic diffractive scattering
amplitude in terms of a colour dipole-dipole scattering amplitude and
the transverse wave functions of the hadrons (or photons) involved
\cite{KraD91, DFK94}.

We follow the notation and normalisation of \cite{RDN}. The scattering
matrix element for the reaction $h_1+h_2 \to h_3 + h_4$ is expressed
in the form
\begin{eqnarray} \label{T}
T &=& 2 i s {{1}\over{4\pi}} \int d^2 R_1 d z_1 {{1}\over{4\pi}} 
\int d^2 R_2 d z_2 \int d^2 b e^{-i \vec b \vec \Delta}
\tilde J(b,\vec R_1,z_1, \vec R_2,z_2)\nonumber\\
&&\times\psi^*_{h3}(\vec R_1,z_1) 
\psi_{h1}(\vec R_1,z_1)\psi^*_{h4}(\vec R_2,z_2) \psi_{h2}(\vec R_2,z_2).
\end{eqnarray}
Here $\psi(\vec R_i, z_i)$ denotes the light-cone wave function of the
hadron $i$ or the hadronic wave function of a photon, $\vec R$ is the
transverse extension, $z$ is the fraction of longitudinal momentum,
$\vec b$ is the impact parameter, $\vec\Delta^2 = t$ is the momentum
transfer and $\tilde J$ is given by eqn.(8) of \cite{RDN}. For a
discussion of these wave functions we refer to \cite{KDP98,DGP98}.

We outline briefly the philosophy behind the choice of the photon wave
function \cite{DGP98}. It is essentially given by the lowest-order
perturbative expression for the quark-antiquark content of the photon,
with chiral symmetry breaking and confinement being simulated by a
$Q^2$-dependent quark mass. This procedure works remarkably well in
quantum-mechanical examples. The quark mass was determined by
comparing the zeroth-order result for the vector-current correlator
with the analytically continued phenomenological expression
($\rho$-pole plus continuum) in the Euclidean region. The resulting
mass is given by
\begin{eqnarray}
m_{u,d} & = & \left\{ \begin{array}{r@{\quad:\quad}l} m_0 \,(1-Q^2/1.05) & Q^2 \le 1.05\\  0& Q^2 \ge 1.05 \end{array} \right. \nonumber\\
m_{s} & = & \left\{ \begin{array}{r@{\quad:\quad}l} 0.15 + 0.16\,(1-Q^2/1.6) & Q^2 \le 1.6\\  0.15 & Q^2 \ge 1.6 \end{array} \right. \nonumber\\
m_c & = & 1.3.
\end{eqnarray}
The parameter $m_0$ for the $u,d$ quarks was found to be $\sim$ 0.22
GeV.

A shortcoming of the stochastic vacuum model applied to high energy
scattering is that it contains no explicit energy dependence. As a
consequence most of the early applications were restricted to a center
of mass energy of 20 GeV. However it is not difficult to insert an
{\em ad hoc} energy dependence, and this is the procedure we adopt
here.

The phenomenological soft pomeron of hadronic interactions cannot
explain deep inelastic scattering or $J/\psi$ photoproduction. A much
stronger energy dependence is observed, and it varies with $Q^2$. In
terms of an effective trajectory the latter fact implies a $Q^2$
dependent intercept \cite{qdep}, in conflict with Regge theory. The
suggestion \cite{shad} that there is a significant amount of shadowing
in soft processes, which disappears at large $Q^2$, so that the
observed effective intercept of $\sim 1.08$ is the result of the
pomeron having a much larger intercept is difficult to sustain. All
soft processes are found to have the same value \cite{pdg}, and
explicit model calculations \cite{Don98} show that the $p\bar p$ total
and differential cross sections restrict the maximum value of the bare
intercept to a much smaller value than is required by the deep
inelastic and $J/\psi$ data.

Consequently a two-pomeron model \cite{DL98} has been proposed, based
on a strict application of the Regge-pole formalism to such processes.
Diffractive processes are described by the exchange of {\em two}
trajectories carrying vacuum quantum numbers with different {\em
  fixed} intercepts and with {\em residues} dependent on $Q^2$. The
latter does not conflict with Regge theory. One of these trajectories
is the conventional non-perturbative soft pomeron with an intercept
$\alpha_{soft} \sim 1.08 $ whose residue is large for hadrons and
photons with low $Q^2$, and a hard pomeron with an intercept
$\alpha_{hard} \sim 1.40$. Before the problems associated with
next-to-leading order contributions to the BFKL pomeron \cite{Gri98},
it would have been tempting to identify this hard pomeron with the
latter, but at present it is simply taken as a phenomenological
prescription to describe the experimental data in a self-consistent
way.

In a recent paper \cite{Rue98} the two pomeron approach scattering of
\cite{DL98} was extended to dipole-dipole scattering. All
dipole-dipole amplitudes where both dipoles are larger than the value
$c = 0.35 $ fm are multiplied by the energy dependent factor:
$(W^2/W_0^2)^{\alpha_{soft}(t) -1}$ with $W_0$ = 20 GeV,
$\alpha_{soft}(t) = 1.08 + 0.25 t$. If one or both dipoles are smaller
than $c$ then the trajectory is replaced by the fixed pole
$\alpha_{hard} = 1.28$. This value of $\alpha_{hard}$ ws chosen as
experimentally $F_2 \propto W^{0.56}$ at $Q^2 = 20$ GeV$^2$ and the
fixed-pole approximation made because of the lack of shrinkage in the
$J/\psi$ photoproduction differential cross section. It turns out that
the model of the stochastic vacuum, which is supposed to be an
approximation to the IR behaviour of QCD, overestimates the
non-perturbative contributions of very small dipoles. Therefore the
non-perturbative contribution is put to zero if one of the dipoles is
smaller than some value $r_{cut}$, which came out as 0.16 fm. A
diffractive scattering amplitude of a (virtual) photon with a hadron
in this model is then given by the general expression
\begin{equation}
T(W^2,Q^2,t) = \beta_{soft}(Q^2,t) (W^2)^{\alpha_{soft}(t)} + \beta_{hard}
(Q^2,t) (W^2)^{\alpha_{hard}(t)}.
\end{equation}

For transverse photons of high virtuality $Q^2$ another source of
energy dependence is induced by the model. Since the hadronic size of
the photon is determined by $1/\sqrt{z (1-z) Q^2}\, , \; z$ being the
longitudinal momentum fraction, there will be at high values of $Q^2$
a large contribution from quarks with very small longitudinal momentum
$z W/2$ or $(1-z)W/2$. The basis of the model is however that the
momentum of the quarks should be large as compared to the Fourier
components of the vacuum fluctuations. Therefore in \cite{Rue98} a
cutoff for $z$ (and $1-z$) was introduced by requiring that $ z W \geq
0.2 $ GeV, $(1-z)W \geq 0.2$ GeV. This cutoff induces an additional
energy dependence at presently accessible energies. Therefore the
intercept of \cite{Rue98} is smaller than that of \cite{DL98}. By
simple Ans\"atze involving only four parameters it was possible to
obtain in this way a good description of data for the proton structure
function and electroproduction of vector mesons, without noticeably
affecting the earlier fits to hadron-hadron scattering. An interesting
and relevant consequence of the model is that the total hadronic cross
section for real photons has a stronger energy dependence than for
purely hadronic scattering as the hard pomeron does not decouple in
the limit $Q^2 \rightarrow 0$. This is a direct result of the hard
component in the photon wave function. It is also a feature of the
two-pomeron model of \cite{DL98}.

The advantage of relating the pomeron-coupling to dipoles (see also
\cite{Nikx}) rather than to hadrons is twofold. First the dipole cross
sections can be calculated directly in the MSV. Secondly once the
coupling scheme is fixed the pomeron residues are determined by the
corresponding wave functions. Thus in principle in the following
predictions of the $\gamma\gamma$ reactions there is {\em no new free
  parameter} involved. However the results for the $\gamma\gamma$
total cross section depend strongly on the quark mass whereas the
reactions we considered previously were quite insensitive to these
parameters. So we allowed for small changes in the quark mass to
obtain the right absolute size of the $\gamma\gamma$ total cross
section. Because of their insensitivity to this parameter these
changes do not significantly affect any of our previous results.

\section{$\gamma\gamma$ and $\gamma^{*}\gamma$ Scattering}

Before making predictions for the $\gamma^{(*)}\gamma^{(*)}
\rightarrow V_1 V_2$ reactions, for which there is currently no high
energy data, we show that the formalism predicts correctly the high
energy data which do exist, namely the total hadronic $\gamma\gamma$
cross section and the photon structure function. In both cases we
restrict ourselves to comparison with the LEP data, as in the former
case the photon energy is sufficiently high, and in the latter case
the values of $x_{Bj}$ are sufficiently small, for our model to be
applicable.

\subsection{The Total Hadronic $\gamma\gamma$ Cross Section}

The total hadronic $\gamma\gamma$ cross section
$\sigma_{\gamma\gamma}$ has been measured recently at LEP by L3
\cite{L3a} and by OPAL \cite{OPALa} for $\gamma\gamma$ center of mass
energies between 5 and 110 GeV. There is a small normalization
difference between the two experiments, but both agree on the energy
dependence which appears stronger than that observed in purely
hadronic interactions. Interpreted in terms of a single pole, the
effective intercept is $1.158 \pm 0.006 \pm 0.028$ \cite{L3a}.

\begin{figure}[ht]
\centering
\begin{minipage}{6.5cm}
\includegraphics[width=6.5cm]{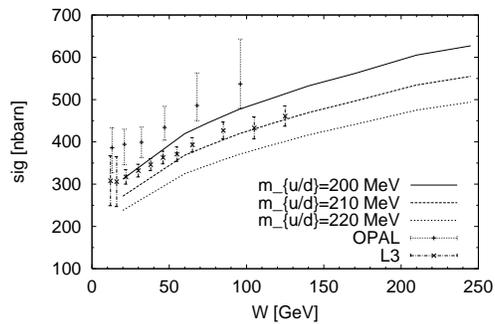}
\end{minipage}
\caption{
  Preliminary experimental data for the total hadronic $\gamma\gamma$
  cross section from L3 \cite{L3a} and OPAL \cite{OPALa} and our
  predictions for different values of the effective quark mass $m_0$
  (see eqn.(3)). The values of $m_0$ are 220 MeV, 210 MeV and 200
  MeV.}
\label{photogammagamma}
\end{figure}

As already discussed the normalisation of $\sigma_{\gamma\gamma}$ is
very sensitive to the quark mass parameter $m_0$. In fact it varies as
$\sim 1/m_0^4$. However the energy dependence is not very sensitive to
$m_0$. The predictions of our model for the the total $\gamma\gamma$
cross section for $m_0$ = 220, 210 and 200 MeV are shown in
Fig.\ref{photogammagamma}. The energy dependence changes very little
across this range of $m_0$, the effective intercept being 1.142,
1.145, 1.149 respectively for the three values. This energy dependence
is a direct consequence of the hard pomeron not decoupling in the $Q^2
\rightarrow 0$ limit of the proton structure function. Here the effect
is more marked as the suppression of the hard component arising from
the coupling to the proton is removed.

A choice of 210 MeV is clearly to be preferred to the initial value of
220 MeV of our previous publications, and we use 210 MeV throughout.
This minor change has no effect on any of the purely hadronic
predictions of the model and actually serves to improve slightly the
description of high energy photon-proton reactions \cite{DGP98,KDP98}.
Our curves fall below the data at smaller values of $W$ because the
non-pomeron Regge contribution, which we have not included, becomes
important there.

\subsection{The Photon Structure Function}

The photon structure function $F_2^\gamma(x,Q^2)$ has been measured
over a wide range of $Q^2$, but primarily at rather large $x$. This
restriction has been due to the machine energies previously available,
but at LEP $F_2^\gamma$ can be measured in the range $x > 10^{-3}$ and
$1 < Q^2 < 10^3$ GeV$^2$ \cite{sr}. These measurements are in the
domain where we expect pomeron dominance, and hence where our model
can make specific predictions. Note that the two-pomeron model does
not factorize simply, so that relating the photon structure function
to the proton structure function $F_2^p$ by \cite{GAL}
\begin{equation}
F_2^\gamma = F_2^p{{\sigma_{\gamma p}(W^2)}\over{\sigma_{p p}(W^2)}}
\end{equation}
is not valid. Obviously it is extremely important to check just how
well simple factorization does hold. Note that the contribution from
the valence quark structure function due to the hadronic content of
the real photon will mask factorization, should it be valid, except at
very small values of $x$.

A reasonable lower limit on the hadronic center-of-mass energy $W$ for
the application of the model is 15 GeV, which we convert to an upper
limit on $x$ for each $Q^2$. It turns out that this is more-or-less
matched to the lower limit on $x$ for which published data exist,
which provides a check on the normalization, at least in principle.
However the data are at values of $x$ at which the valence quarks from
the hadronic component of the photon contribute, and an estimate has
to be made of this. Our direct predictions for the photon structure
function, together with the data, are shown in Fig.2 by the dashed
line. These are parameter free predictions and determine the photon
structure function at small $x$. In view of possible forthcoming LEP
data we have extended the calculations to values of $x$ beyond those
for which data are currently available.

As anticipated, our predictions are low in the region of overlap with
current data as there is still a significant contribution from the
valence quark structure function of the hadronic content of the real
photon ($\rho$, $\omega$, $\phi$ etc.) \cite{sas,grv1,afg}. In naive
vector meson dominance (VMD) this is given by
\begin{equation}
{{1}\over{\alpha}}F_{2,had}^\gamma(x,Q^2) = F_{val}^\pi(x,Q^2)
\sum_V{{4\pi}\over{f_V^2}},
\end{equation}
where the sum is usually over $\rho$, $\omega$, $\phi$. The additional
assumption has been made that the vector meson structure functions can
all be represented by the valence structure function of the pion
$F_{val}^\pi(x,Q^2)$. This in itself is quite an extreme statement, as
there is no obvious reason why the structure function of the
short-lived vector mesons should be the same as those of the
long-lived pion. Additionally it is not clear whether one should take
the simple incoherent sum or allow for coherence effects. Finally,
higher mass vector mesons must also make some contribution, but this
is almost certainly small compared to the uncertainties in any
estimate of the hadronic component.

To add to these uncertainties, the pion structure function is only
known experimentally for $x > 0.2$. To obtain the structure function
in the kinematical domain of interest here, it is necessary to use the
DGLAP evolution equations to fit the data and to extrapolate
\cite{aur,grv2}. This was the approach used in fitting $F_2^\gamma$ by
\cite{grv1,afg}, although with somewhat different assumptions about
the effective strength of the contribution. In contrast, in \cite{sas}
the {\em shape} of the hadronic contribution was left free to be
determined by the data, but the {\em normalization} was fixed.

\begin{figure}[ht]
\leavevmode
\centering
\begin{minipage}{6.5cm}
\includegraphics[width=6.5cm]{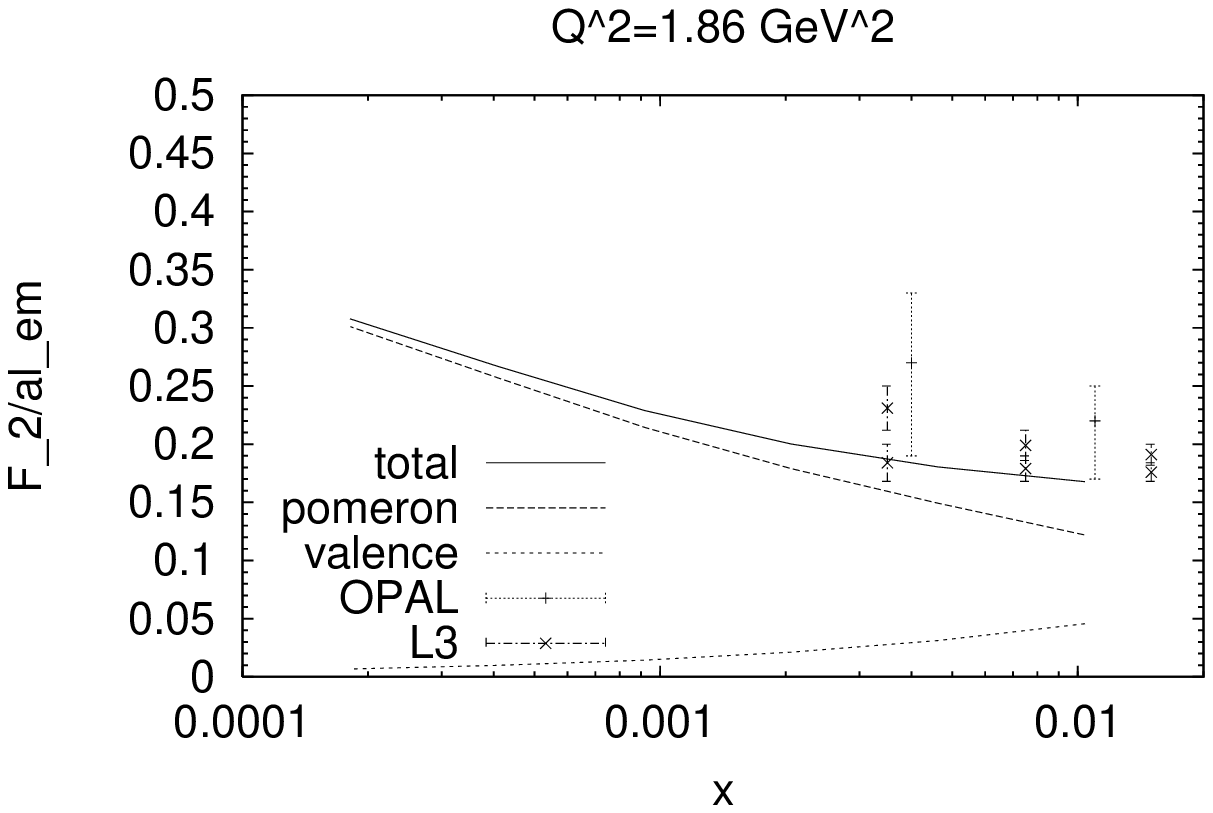}
\end{minipage}
\hfill
\begin{minipage}{6.5cm}
\includegraphics[width=6.5cm]{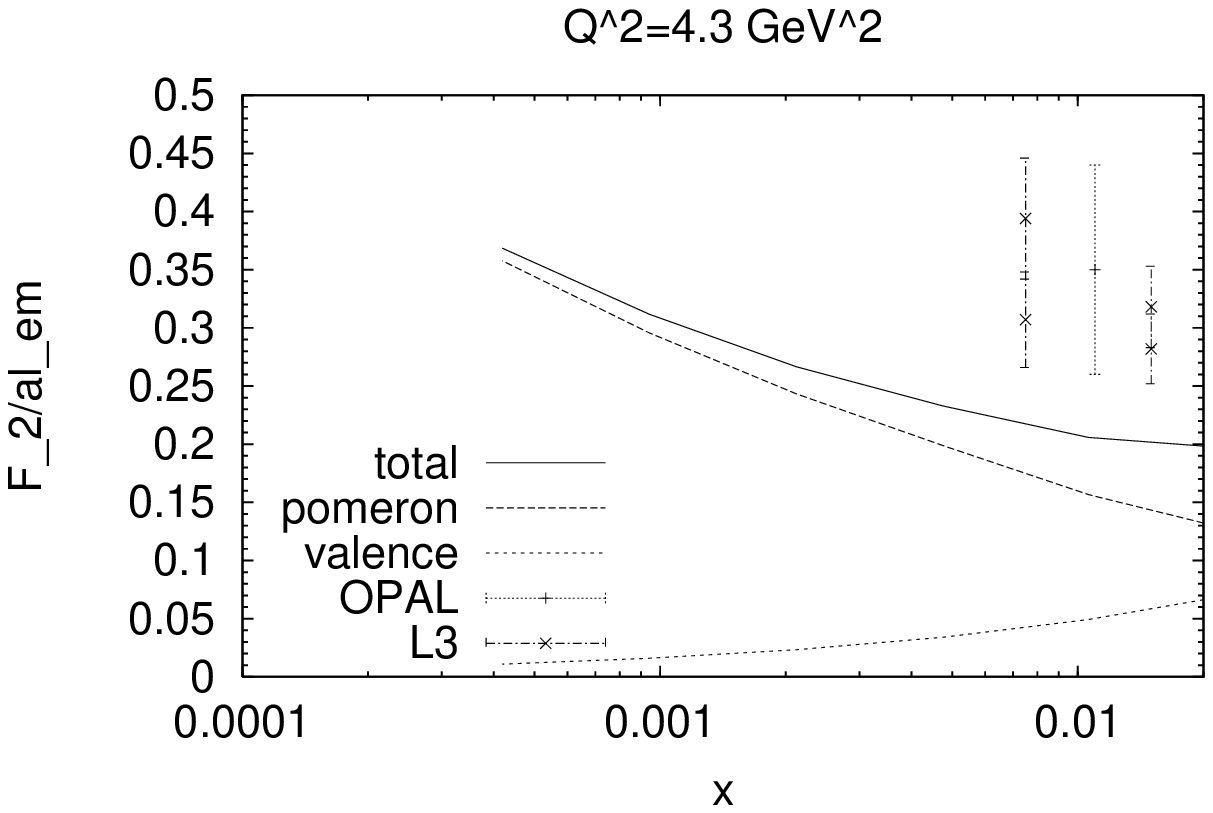}
\end{minipage}
\begin{minipage}{6.5cm}
\includegraphics[width=6.5cm]{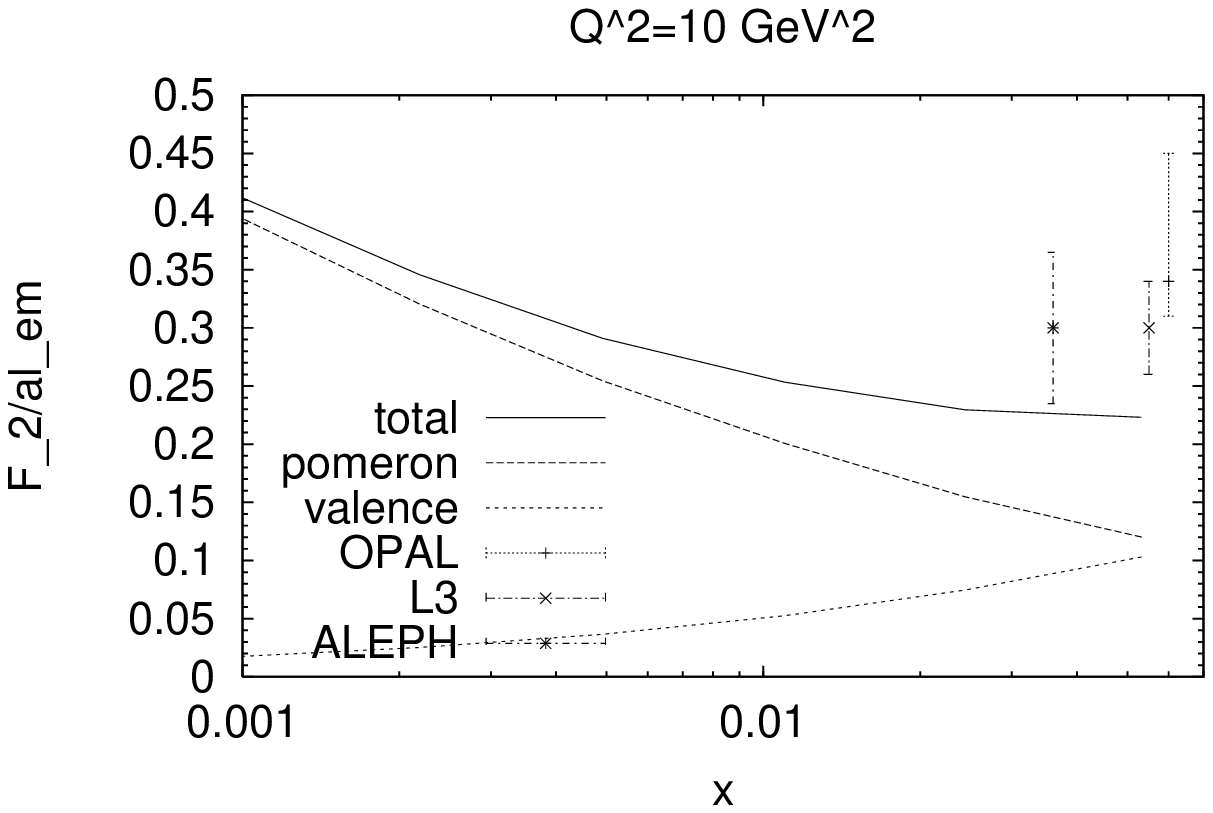}
\end{minipage}
\hfill
\begin{minipage}{6.5cm}
\includegraphics[width=6.5cm]{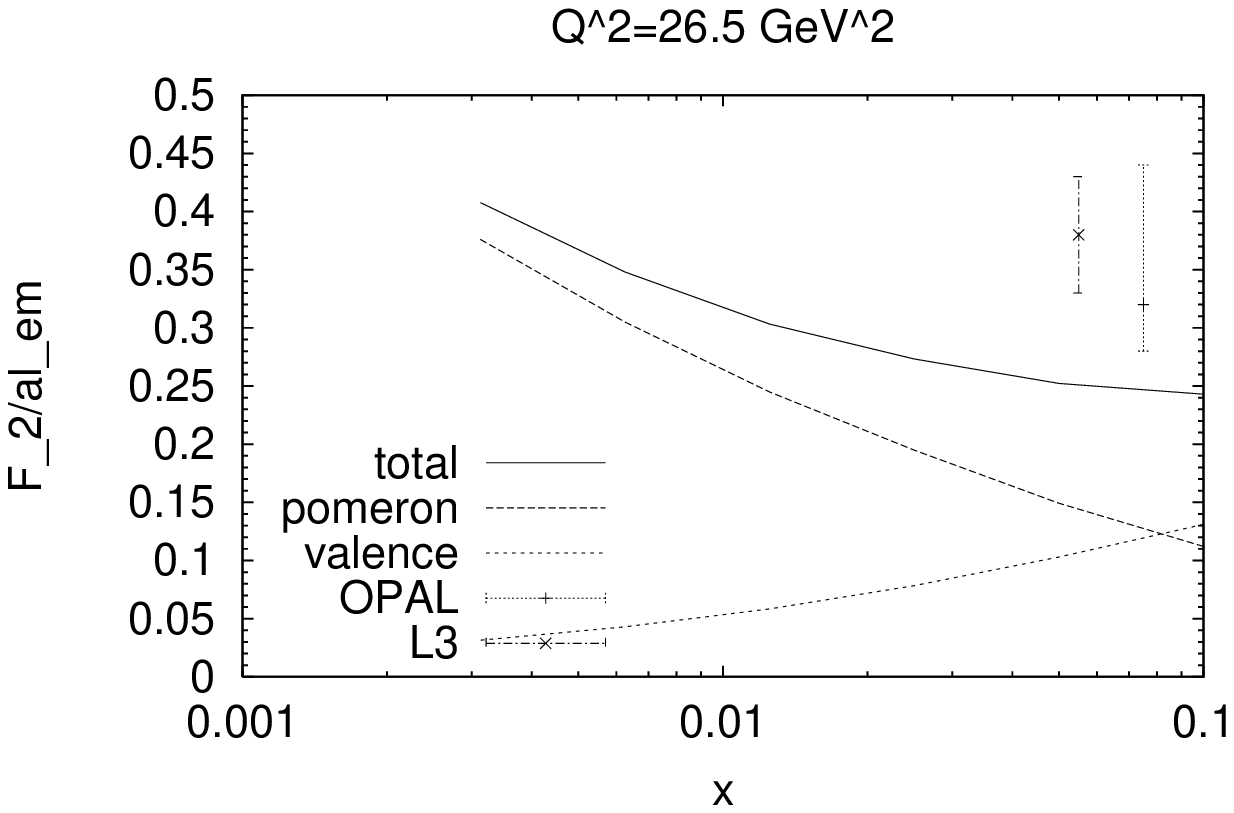}
\end{minipage}
\caption{
  Data and predictions for the structure function
  $F_2^{\gamma}/\alpha_{em}$ at small $x$. The data comprise
  preliminary data from Aleph \cite{ALEPH} and L3 \cite{L3b} and
  published data from OPAL \cite{OPALb}. The dashed line is the
  sea-quark (pomeron exchange) contribution, which dominates for small
  $x$, calculated in our approach without any free parameters. The
  dotted line is the estimated valence quark contribution (see text)
  and the solid line the sum of the two. The maximum $x$-value is
  chosen to give $\surd s \ge 15$ GeV. The mean values of $Q^2$ are
  1.86 GeV$^2$, 4.3 GeV$^2$, 10.0 GeV$^2$ and 26.5 GeV$^2$.}
\label{F2}
\end{figure}

For definiteness we have used the DGLAP evolved pion structure
function of \cite{grv1}, and have retained only the $\rho$, $\omega$
and $\phi$ in the sum of eqn.(6). Combining this with the predictions
of our model for the singlet term gives a good description of the
small-$x$ structure function. A comparison with the data of
\cite{ALEPH,L3b,OPALb} is given in Fig.\ref{F2}.

At smaller values of $x$, where pomeron exchange dominates and the
valence quark contribution declines with increasing $Q^2$, our
predictions for $F_2^\gamma$ should be sufficient on their own.  They
agree rather well with the simple factorisation formula of eqn.(5) at
small $Q^2$, but increasingly diverge as $Q^2$ increases. This is not
unexpected, as at small $Q^2$ the proton structure function is
dominated by a single term \cite{DL98,Rue98} and factorization is a
reasonable approximation. Interestingly at higher $Q^2$ the
predictions match well to those of \cite{sas}, but lie below them at
the lower values of $Q^2$.

\section{$\gamma^{(*)}\gamma^{(*)} \rightarrow V_1 V_2$}

The model allows calculations to be performed for any vectors $V_1$,
$V_2$ in the light-quark or heavy-quark sectors, including radial
excitations, and for any virtuality of the two photons. For simplicity
we restrict ourselves to the following representative cases:
$\gamma^*(Q_1^2)\gamma^*(Q_2^2) \rightarrow \rho^0\rho^0$ for $0 \le
Q_1^2 \le 10$ GeV$^2$, $0 \le Q_2^2 \le 10$ GeV$^2$,
$\gamma^*(Q_1^2)\gamma\rightarrow \rho^0\phi$ and
$\gamma^*(Q_1^2)\gamma\rightarrow \rho^0 J/\psi$ for $Q_1^2=0$ and 2.5
GeV$^2$ and $\gamma\gamma \rightarrow J/\psi J/\psi$ for real photons.
The latter is dominated by pomeron exchange at all energies and the
former at sufficiently high energy. We note that the dominance of
pomeron exchange can be assured in the light-quark sector at any
energy by considering $\gamma^{(*)}\gamma^{(*)} \rightarrow \phi\phi$
or $\gamma^{(*)}\gamma^{(*)} \rightarrow \rho\phi$, although the cross
sections are obviously appreciably smaller than those for $\rho\rho$
production.

\begin{figure}[ht]
\leavevmode
\centering
\begin{minipage}{6.5cm}
\includegraphics[width=6.5cm]{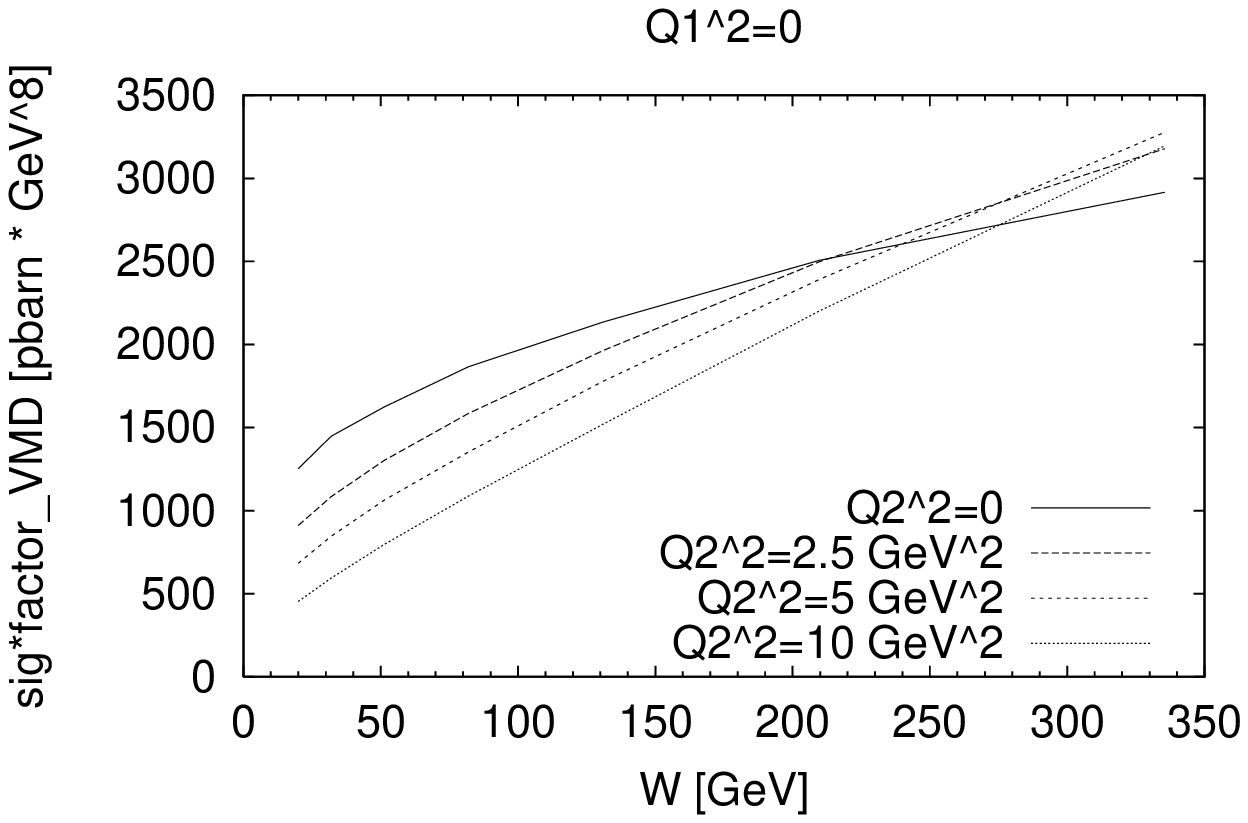}
\end{minipage}
\hfill
\begin{minipage}{6.5cm}
\includegraphics[width=6.5cm]{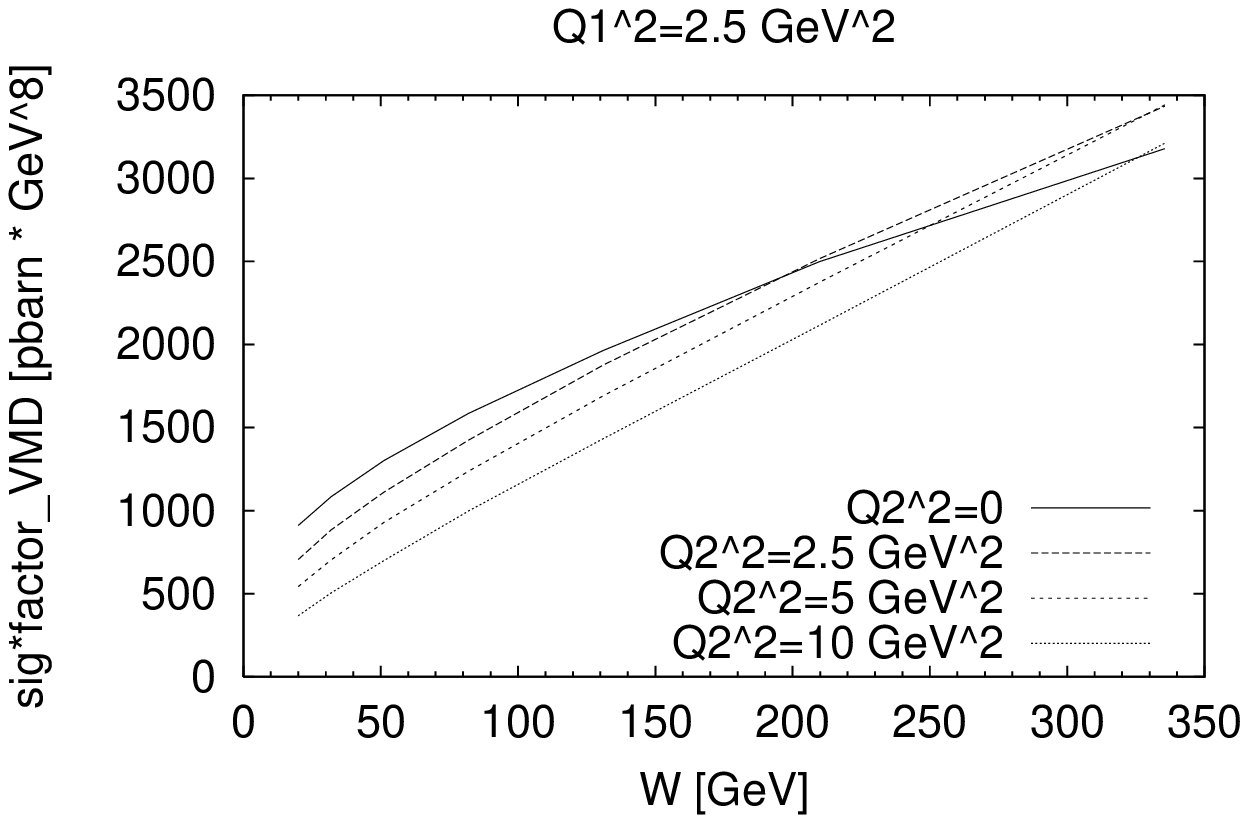}
\end{minipage}
\begin{minipage}{6.5cm}
\includegraphics[width=6.5cm]{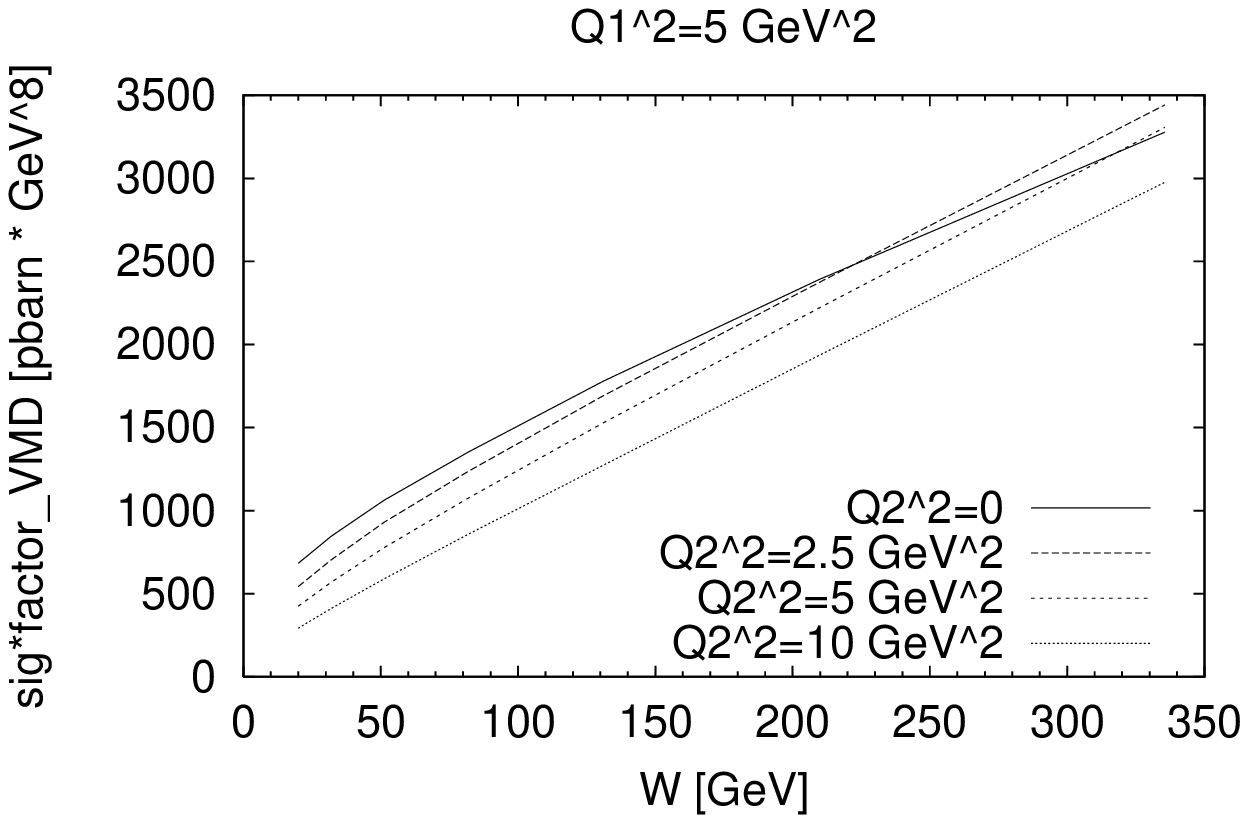}
\end{minipage}
\hfill
\begin{minipage}{6.5cm}
\includegraphics[width=6.5cm]{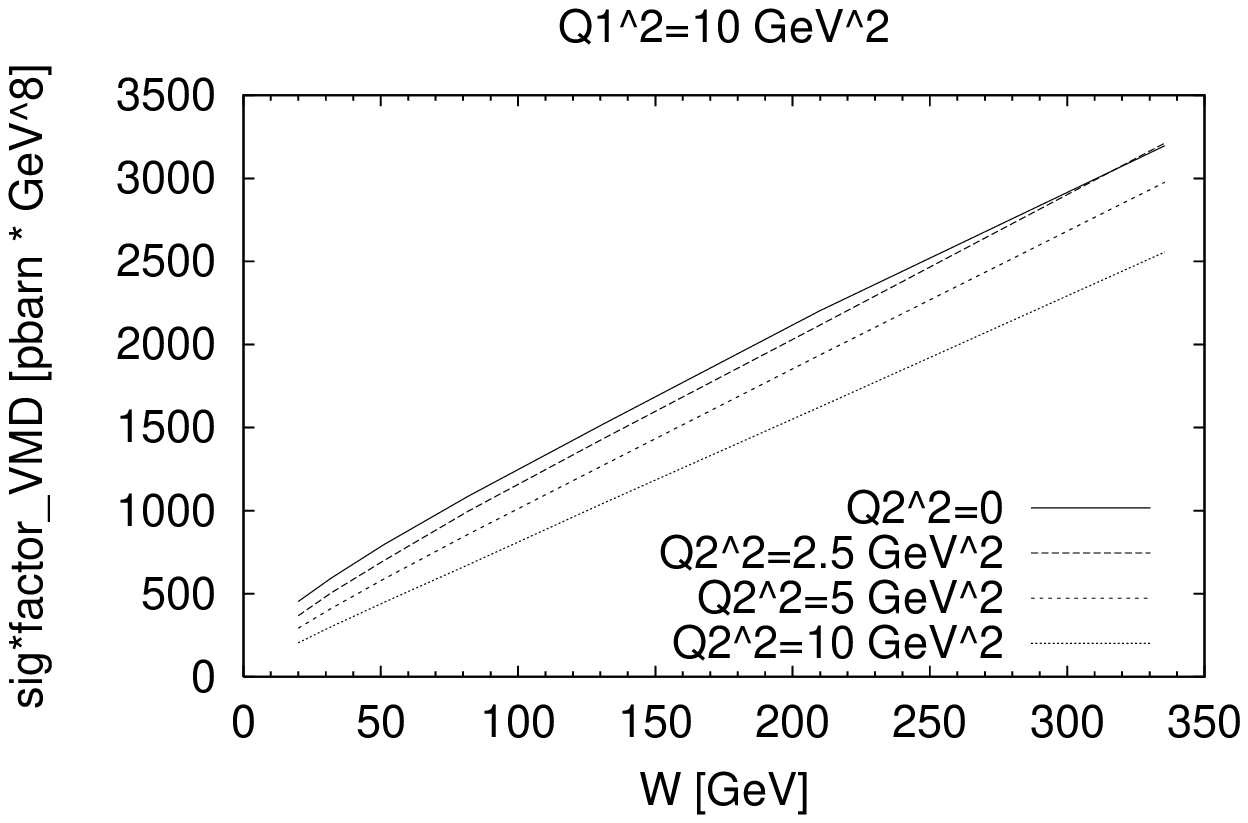}
\end{minipage}
\caption{
  Predicted cross sections for the reaction $\gamma^{(*)}\gamma^{(*)}
  \rightarrow \rho\rho$. The virtuality of the first photon is $Q_1^2
  = 0$ GeV$^2$, 2.5 GeV$^2$, 5.0 GeV$^2$ and 10 GeV$^2$. In each case
  the same virtualities are shown for the second photon. The cross
  sections have been scaled by the VMD factor
  $(Q_1^2+m_{\rho}^2)^2(Q_2^2+m_{\rho}^2)^2$.}
\label{RhoRho}
\end{figure}

The results for the $\rho\rho$ total cross section are shown in
Fig.\ref{RhoRho} as a function of the $\gamma^*\gamma^*$
center-of-mass energy $W$, multipled by the naive VMD factor
$(Q_1^2+m_\rho^2)^2(Q_2^2+m_\rho^2)^2$.  This is to provide a common
scale, and to demonstrate quite explicitly that the cross sections do
{\em not} follow simple VMD. That they do not is unsurprising. The
effect of the hard pomeron contribution is explicitly {\em not} like
VMD, and the $Q^2$-dependence is more intricate than simple VMD as it
depends on the size of the dipole. The other notable difference is
that the energy dependence changes with increasing $Q^2$ as opposed to
the $Q^2$-independent behaviour expected of simple VMD.  The results
show clearly the dominance of the soft pomeron for small photon
virtuality and the increasing importance of the hard pomeron as the
photon virtuality increases. The dominance of the hard pomeron term
sets in much more rapidly than in $\gamma^* p$ reactions. Its presence
is already visible at $Q^2 = 0$ and by $Q^2 = 2.5$ GeV$^2$ it is
dominating the cross section.

\begin{figure}[ht]
\leavevmode
\centering
\begin{minipage}{6.5cm}
\includegraphics[width=6.5cm]{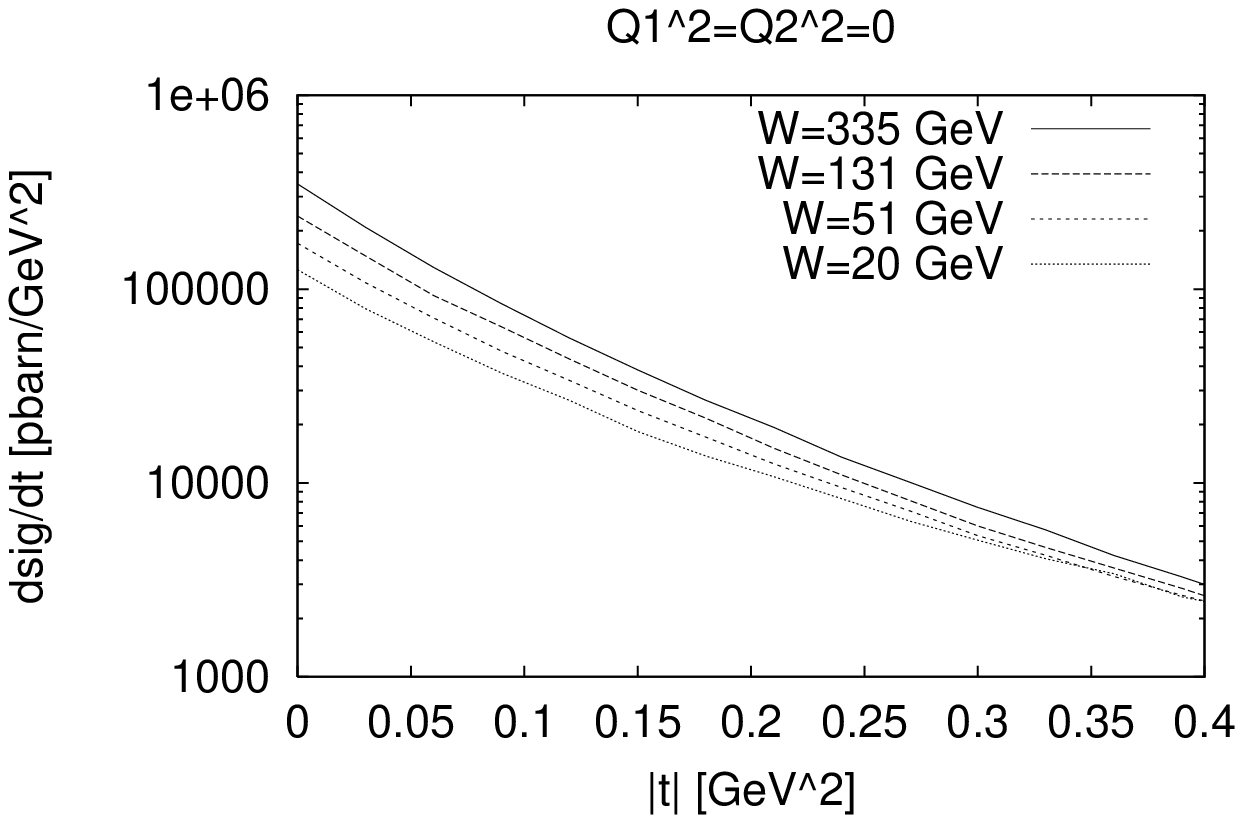}
\end{minipage}
\hfill
\begin{minipage}{6.5cm}
\includegraphics[width=6.5cm]{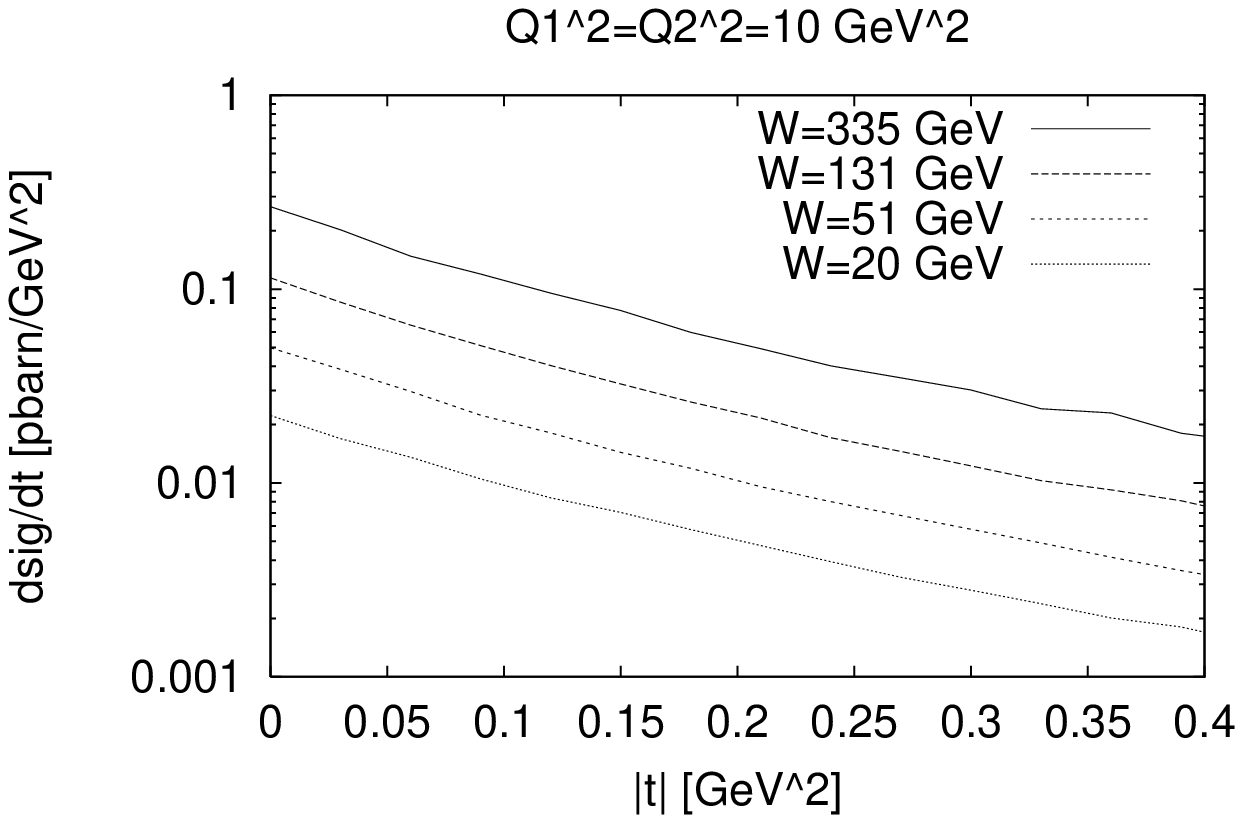}
\end{minipage}
\caption{
  The differential cross section $d\sigma/dt$ for $\gamma\gamma
  \rightarrow \rho\rho$ for $Q_1^2 = Q_2^2 = 0$ and $Q_1^2 = Q_2^2
  =10$ GeV$^2$. These cross sections have {\em not} been scaled by a
  VMD factor.}
\label{RhoRhodsigdt}
\end{figure}

The differential cross sections $d\sigma/dt$ ({\em without} the VMD
normalization factor) are shown in Fig.\ref{RhoRhodsigdt} for the two
extremes $Q_1^2 = Q_2^2 = 0$ and $Q_1^2 = Q_2^2 = 10$ GeV$^2$.
Shrinkage, that is an energy dependence of the slope of $d\sigma/dt$,
due to the Regge factor $2\alpha'(t) \log s$ is clearly visible for
the case of two real photons, emphasizing the dominant contribution
from the soft pomeron. Shrinkage is much less obvious for the case of
large photon virtuality, where the hard pomeron is now the dominant
contribution. Figure 4 also shows the decrease in cross section with
increasing photon virtuality which is masked in Fig.3 by the VMD
normalization factor.

\begin{figure}[ht]
\leavevmode
\centering
\begin{minipage}{6.5cm}
\includegraphics[width=6.5cm]{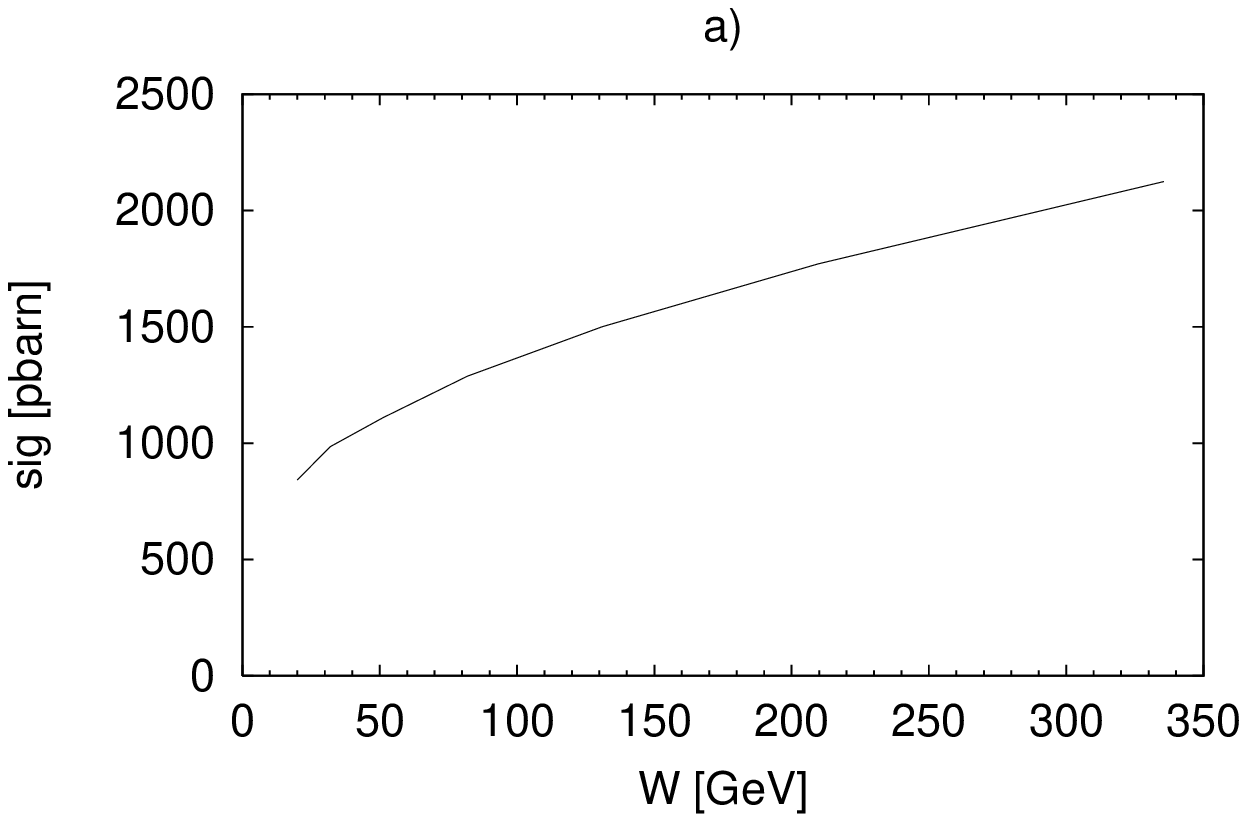}
\end{minipage}
\hfill
\begin{minipage}{6.5cm}
\includegraphics[width=6.5cm]{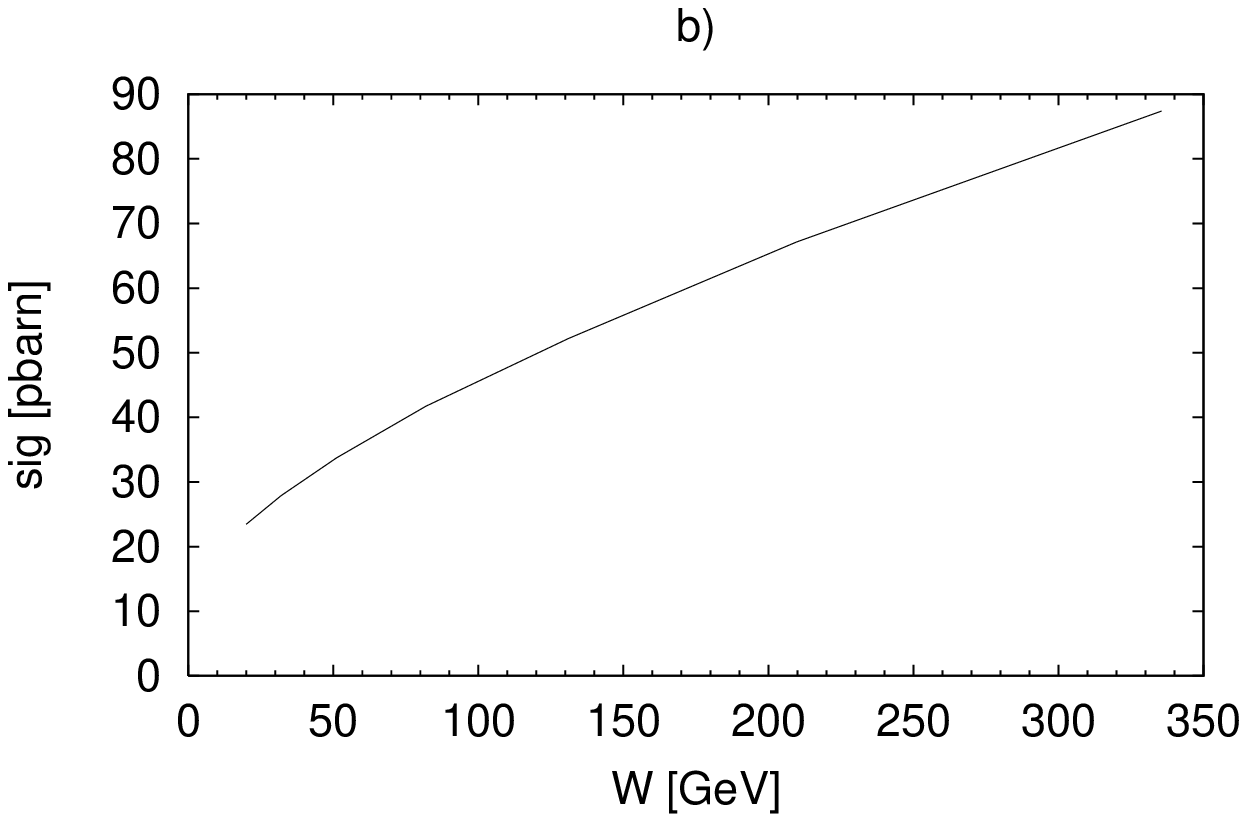}
\end{minipage}
\begin{minipage}{6.5cm}
\includegraphics[width=6.5cm]{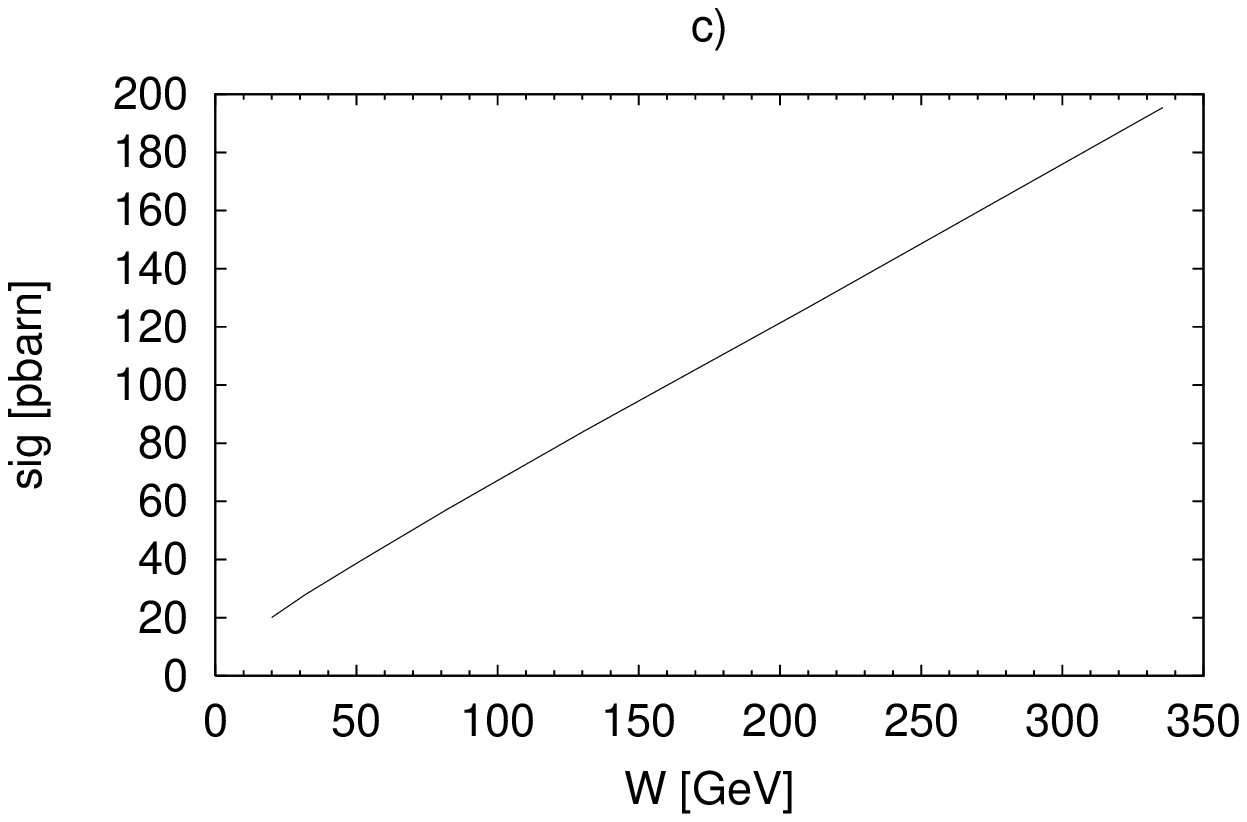}
\end{minipage}
\hfill
\begin{minipage}{6.5cm}
\includegraphics[width=6.5cm]{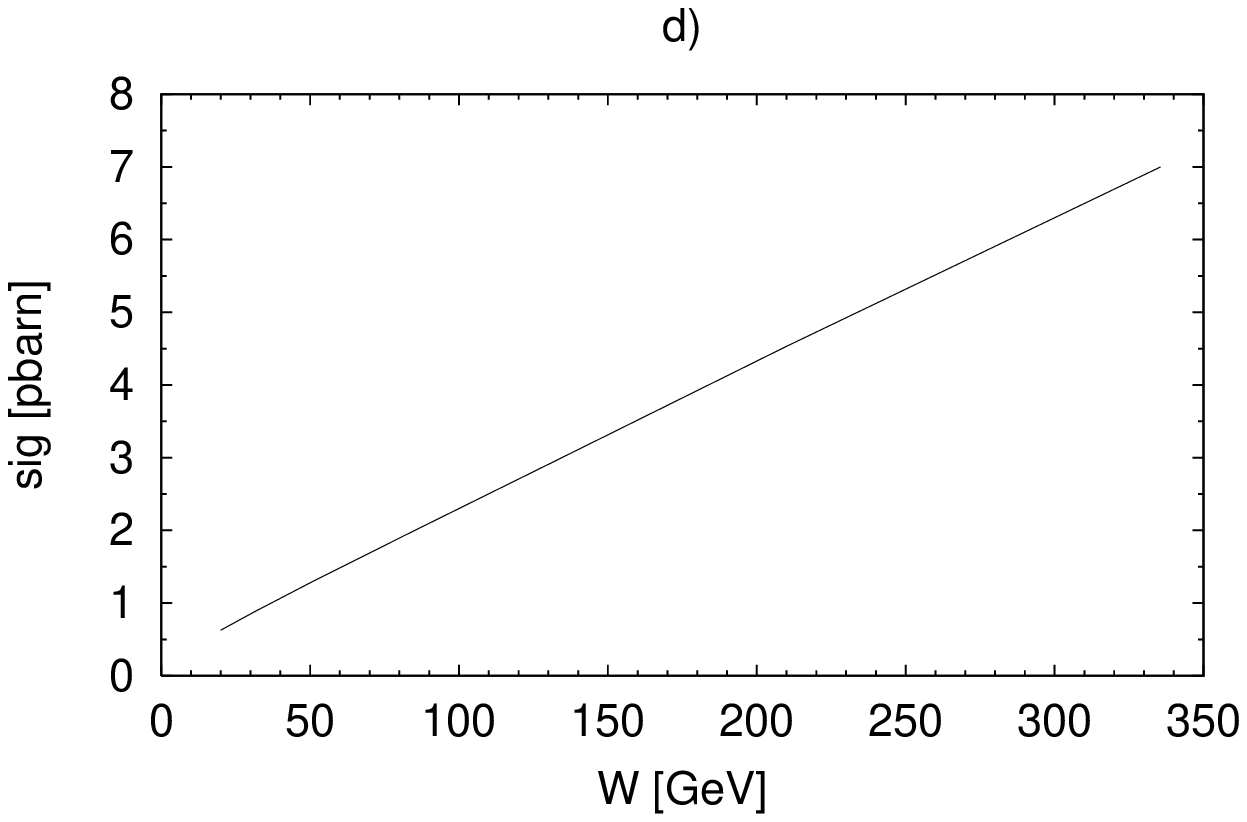}
\end{minipage}
\begin{minipage}{6.5cm}
\includegraphics[width=6.5cm]{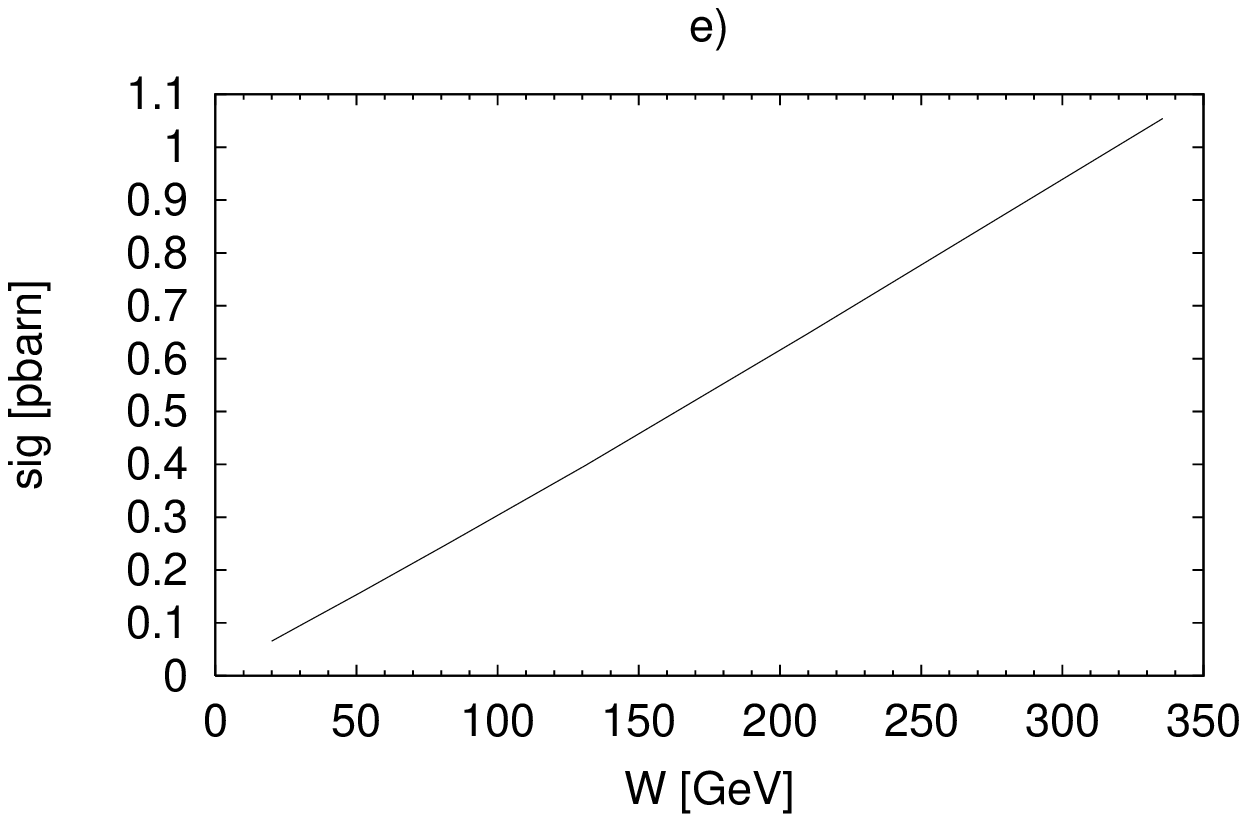}
\end{minipage}
\caption{
  Total cross section of vector meson production in photon-photon
  scattering as a function of energy: a) $\gamma\gamma \rightarrow
  \rho \phi$, b) $\gamma^*\gamma \rightarrow \rho \phi$ for
  \mbox{$Q_1^2=2.5$ GeV$^2$}, c) $\gamma\gamma \rightarrow \rho
  J/\psi$, d) $\gamma^*\gamma \rightarrow \rho J/\psi$ for
  \mbox{$Q_1^2=$ 2.5 GeV$^2$} and e) $\gamma\gamma \rightarrow J/\psi
  J/\psi$.}
\label{JPsiJPsi}
\end{figure}

In Fig.\ref{JPsiJPsi} we show the cross sections for $\gamma^*\gamma
\rightarrow \rho \phi$ and $\gamma^*\gamma \rightarrow \rho J/\psi$ to
illustrate that even for quite modest photon virtuality on the $\rho$
the hard pomeron dominates. The total cross section for $\gamma\gamma
\rightarrow J/\psi J/\psi$ is also shown in Fig.\ref{JPsiJPsi}. As it
is already very small, we do not give any results for non-zero photon
virtuality. In practice the measurable cross section will be much less
as we have taken no account of the branching fraction of the $J/\psi$
to $\mu^+\mu^-$. The almost complete dominance of the hard pomeron is
obvious, and is more marked than for $J/\psi$ photoproduction.

\section{Conclusions}

We have demonstrated that the two-pomeron modification \cite{Rue98} of
the MSV model correctly predicts the energy dependence of the
$\gamma\gamma$ total cross section without any adjustment of
parameters. However to obtain the correct normalization it is
necessary to change the quark mass to 210 MeV from the 220 MeV of our
previous publications. This then gives good agreement for $W>50$ GeV
where we can neglect the non-pomeron Regge contributions. Our results
also agree with the photon structure function at small $x$, within the
limitations of the present data and the uncertainties on the valence
quark content of the hadronic component of the photon. This justifies
the use of the model to estimate cross sections for
$\gamma^{(*)}\gamma^{(*)} \rightarrow V_1V_2$. These are found to be
at a level which, in principle, allows a rather direct determination
of the colour dipole-dipole cross section. Similar results for the
case of purely real photons have been obtained recently in a rather
different model \cite{block}. We stress the importance of photon
virtuality, even if small, in probing the most interesting part of
high-energy scattering.

\begin{figure}[ht]
\centering
\begin{minipage}{6.5cm}
\includegraphics[width=6.5cm]{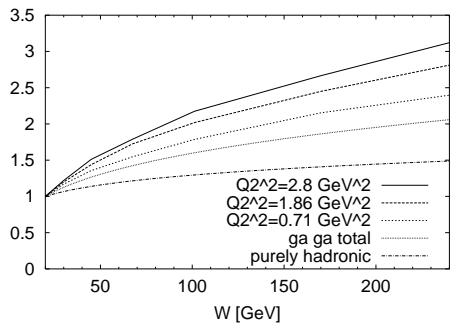}
\end{minipage}
\caption{
  Comparison of the energy dependence of a typical purely hadronic
  total cross section, the $\gamma\gamma$ total cross section and the
  $\gamma\gamma^*$ total cross section for $Q_2^2 = $0.71 GeV$^2$,
  1.86 GeV$^2$ and 2.8 GeV$^2$ to illustrate how rapidly the hard
  pomeron term dominates in $\gamma\gamma^*$ scattering. Because here
  we are only interested in the energy dependence we scaled all cross
  sections to the same value at $W=20$ GeV.}
\label{sigtotbehavior}
\end{figure}

Apart from taking into account a hard pomeron our choice of the photon
wave function is decisive for the results. The hard part of it leads
to a non-negligible coupling of the hard pomeron even to the real
photon, in contrast with purely hadronic reactions. This hard
component, already present in the real photon, and the small value of
the mass parameter $m_0 = 210$ MeV, also explains why the hard
contribution plays a decisive role even at moderate virtualities. We
demonstrate these features graphically in Fig.\ref{sigtotbehavior}.

\section*{Acknowledgments}
The authors thank Peter Landshoff and Otto Nachtmann for interesting
discussions. One of us (A.D.) wants to thank the Institut f\"ur
Theoretische Physik der Universit\"at Heidelberg for the hospitality
extended to him during his stay there and the Bundesministerium f\"ur
Bildung, Wissenschaft, Forschung und Technologie for financial
support. H.G.D. acknowledges support from DAAD and M.R.~is grateful to
the MINERVA-Stiftung for financial support.


\begin{thebibliography}{99}

\bibitem{Nik91}
N.N. Nikolaev and B.G. Zacharov: Z.Phys. {\bf C49} (1991) 607

\bibitem{KraD91}
A. Kr\"amer and H.G. Dosch: Phys.Lett. {\bf B272} (1991) 114

\bibitem{Dos87} 
H.G. Dosch: Phys.Lett. {\bf B190} (1987) 555

\bibitem{DosS88} 
H.G. Dosch and Yu.A. Simonov: Phys.Lett. {\bf B205} (1988) 339

\bibitem{Nac91}
O. Nachtman: Ann.Phys. {\bf 209} (1991) 436

\bibitem{DFK94} 
H.G. Dosch, E. Ferreira and A. Kr\"amer: Phys.Rev. {\bf D50} (1994) 1992

\bibitem{BerN98}
E. Berger and O. Nachtmann: hep-ph/9808320, Eur.Phys.J. {\bf C} (1999) DOI 10.1007/s100529801026

\bibitem{DGKP97}
H.G. Dosch, T. Gousset, G. Kulzinger and H.J. Pirner: Phys.Rev. {\bf D55}
(1997) 2602

\bibitem{KDP98}
G. Kulzinger, H.G. Dosch and H.J. Pirner: hep-ph/9806352, Eur.Phys.J. {\bf C} (1999) DOI 10.1007/s100529800986

\bibitem{Rue98}
M. Rueter: hep-ph/9807448, Eur.Phys.J. {\bf C} (1999) DOI 10.1007/s100529801002

\bibitem{Nikx}
N.N. Nikolaev and B.G. Zakharov: Phys.Lett. {\bf B327} (1994) 149; ibid 157\\
J. Nemchik, N.N. Nikolaev and B.G. Zakharov: Phys.Lett. {\bf B341} (1994) 228

\bibitem{nem}
J. Nemchik, N.N. Nikolaev, E. Predazzi and B.G. Zacharov: Phys.Lett.
{\bf B374} (1996) 199 and Z.Phys. {\bf C75} (1997) 71

\bibitem{dr}
M. Rueter and H.G. Dosch: Phys.Rev. {\bf D57} (1998) 4097

\bibitem{muel}
A.H. Mueller: Nucl.Phys. {\bf B415} (1994) 373\\
A.H. Mueller and H. Patel: Nucl.Phys. {\bf B425} (1994) 471

\bibitem{DL98} A. Donnachie and P.V. Landshoff: Phys.Lett. {\bf B437} (1998) 408

\bibitem{RDN}
M. Rueter, H.G. Dosch and O.Nachtmann: hep-ph/9806342, Phys.Rev. {\bf D59} (1999) 014018

\bibitem{DGP98}
H.G. Dosch, T. Gousset and H.J. Pirner: Phys.Rev. {\bf D57} (1998) 1666

\bibitem{qdep}
A. Capella, A. Kaidalov, C. Merino and J. tran Thanh Van: Phys.Lett.  
{\bf B337} (1994) 358\\
M. Bertini, M. Giffon and E. Predazzi: Phys.Lett. {\bf B349} (1995) 561 

\bibitem{shad}
E. Gotsman, E.M. Levin and U. Maor: Phys.Rev. {\bf D49} (1994) 4321

\bibitem{pdg}
Particle Data Group: Eur.Phys.J. {\bf 3} (1998) 205 

\bibitem{Don98}
S.V. Goloskokov, S.P. Kuleshov and O.V. Selyugin: hep-ph/9409383\\
A. Donnachie: private communication

\bibitem{Gri98}
V.S. Fadin and L.N. Lipatov: Phys.Lett. {\bf B429} (1998) 127

\bibitem{L3a}
L3 Collaboration: M. Acciari et al: Phys.Lett. {\bf B408} (1997) 450\\
L3 Collaboration: {\em Cross section of hadron production in $\gamma \gamma$ collisions at LEP}, talk at {\em XXIX ICHEP}, Vancouver, Canada, 1998

\bibitem{OPALa}
OPAL Collaboration: F. W\"ackerle: talk at {\em XXVIII Int.~Symp.~on 
Multiparticle Dynamics}, Frascati, Italy, 1997

\bibitem{sr}
S. S\"oldner-Rembold: talk at {\em XVIII ISLEPHI}, Hamburg, Germany, 1997 

\bibitem{GAL}
H. Abramowicz, E. Gurvich and A. Levy: Phys.Lett. {\bf B420} (1998) 104

\bibitem{afg}
P. Aurenche, M. Fontannaz and J.Ph. Guillet: Z.Phys. {\bf C64} (1994) 621

\bibitem{grv1}
M. Gl\"uck, E. Reya and A. Vogt: Phys.Rev. {\bf D46} (1992) 1973

\bibitem{sas}
G.A. Schuler and T. Sj\"ostrand: Z.Phys. {\bf C68} (1995) 607

\bibitem{aur}
P. Aurenche: Phys.Lett. {\bf B233} (1989) 517

\bibitem{grv2}
M. Gl\"uck, E. Reya and A. Vogt: Z.Phys. {\bf C53} (1992) 651

\bibitem{ALEPH} 
Aleph Collaboration: {\em Measurement of the Photon Structure Function 
at $Q^2$ of 8.9 and 19.1 GeV$^2$}, talk at {\em XXIX ICHEP}, Vancouver, Canada, 1998

\bibitem{L3b} 
L3 Collaboration: CERN-EP/98-98 and {\em Study of the Hadronic 
Photon Structure Function $F_2^{\gamma}$ at $\surd s \sim $ 183 GeV}, talk at {\em XXIX ICHEP}, Vancouver, Canada, 1998

\bibitem{OPALb} 
OPAL Collaboration: Phys.Lett. {\bf B411} (1997) 387 and
Phys.Lett. {\bf B412} (1997) 225

\bibitem{block}
M.M. Block, E.M. Gregores, F. Halzen and G. Pancheri:\\hep-ph/9809403

\end{thebibliography}
\end{document}